# Synergistic effect of work function and acoustic impedance mismatch for improved thermoelectric performance in GeTe/WC composite

Ashutosh Kumar[a,*], Preeti Bhumla[b], Artur Kosonowski[a], Karol Wolski[c], Szczepan Zapotoczny[c], Saswata Bhattacharya[b,*] Krzysztof T. Wojciechowski[a,*]

[a]Faculty of Materials Science and Ceramics, AGH University of Science and Technology, Kraków 30-059, Poland
[b]Department of Physics, Indian Institute of Technology Delhi, New Delhi 110016, India
[c]Faculty of Chemistry, Jagiellonian University, Gronostajowa 2, Krakow 30-387, Poland

**Abstract:** The preparation of composite materials is promising for concurrent optimization of electrical and thermal transport properties to realize an improved thermoelectric (TE) performance. We report the effect of work function and acoustic impedance mismatch (AIM) on the TE properties of (1-$z$)Ge$_{0.87}$Mn$_{0.05}$Sb$_{0.08}$Te/($z$)WC composite. In particular, a composite consisting of Mn and Sb co-doped GeTe as a matrix and WC (tungsten carbide) as a dispersed phase is prepared, and its structural and TE properties are investigated. The simultaneous increase in the electrical conductivity ($\sigma$) and Seebeck coefficient ($\alpha$) with WC volume fraction ($z$) results in an enhanced power factor ($\alpha^2\sigma$) in the composite. The rise in $\sigma$ is attributed to increased carrier mobility in the composite. This is further established from the work function measurement using the Kelvin probe force microscopy (KPFM) technique and is also supported by the density functional theory (DFT) calculations. The difference in elastic properties (sound velocity) between Ge$_{0.87}$Mn$_{0.05}$Sb$_{0.08}$Te and WC results in a high AIM that leads to a large interface thermal resistance ($R_{int}$) between the phases. The correlation between $R_{int}$ and the Kapitza radius results in reduced phonon thermal conductivity ($\kappa_{ph}$) of the composite and is discussed using the Bruggeman asymmetrical model. The decrease in $\kappa_{ph}$ is further established using phonon dispersion calculations that indicates the decrease in phonon group velocity in the composite. The simultaneous effect of enhanced $\alpha^2\sigma$ and reduced $\kappa_{ph}$ results in a maximum figure of merit ($zT$) of 1.93 at 773K for (1-$z$)Ge$_{0.87}$Mn$_{0.05}$Sb$_{0.08}$Te/($z$)WC composite having $z$=0.010. It results in an average $zT$ ($zT_{av}$) of 1.02 for a temperature difference ($\Delta$T) of 473 K. This study shows promise to achieve higher $zT_{av}$ across a wide range of composite materials having similar electronic structure and different elastic properties.

*corresponding author(s):
Ashutosh Kumar, Email: science.ashutosh@gmail.com
Saswata Bhattacharya, Email: saswata@physics.iitd.ac.in
Krzysztof T. Wojciechowski, Email: wojciech@agh.edu.pl

## 1. Introduction

The majority of energy generated from several energy sources is wasted during practical applications. Therefore a waste heat recovery technology is essential for the efficient use of these energy resources. Thermoelectric (TE) energy conversion technology that can harvest waste heat energy using temperature gradient without emitting pollution is a propitious solution for waste heat recovery and niche power resources.[1,2] The usefulness of TE materials depends on its figure of merit ($zT$), defined as

$$zT = \frac{\alpha^2 \sigma}{(\kappa = \kappa_{ph} + \kappa_e)} T \qquad (1)$$

where $T$ is absolute temperature, $\alpha^2\sigma$ is known as power factor that includes Seebeck coefficient ($\alpha$) and electrical conductivity ($\sigma$), and $\kappa$ is total thermal conductivity consisting of electronic thermal conductivity ($\kappa_e$) and phonon thermal conductivity ($\kappa_{ph}$). The strong coupling between $\sigma$, $\alpha$, and $\kappa_e$ with trade-off relationship is challenging to achieve a high zT in a single system.[3,4] The electrical transport has been optimized through several innovative transport mechanisms to achieve enhanced $\alpha^2\sigma$ in single-phase materials.[5–7] Further, the reduction in $\kappa_{ph}$ has been exemplified in literature through several strategies that amplify the phonon scattering, including lattice defects[8,9], artificial superlattices[10], mass disorder[11], nanostructuring[12], preparation of composite materials.[13,14] However, scattering of phonons in TE materials *via* impurities, lattice anharmonicity, and defects also reduces $\sigma$ due to the scattering of charge carriers.[3]

Moreover, the preparation of composite materials is promising for concurrent optimization of electrical and thermal transport properties to realize an improved TE performance.[15–20] The simultaneous effect of carrier energy filtering and enhanced phonon scattering at the interface between $Bi_{0.4}Sb_{1.6}Te_3/Cu_2Se$ nanocomposite results in an enhanced zT (~1.6 at 488 K).[16] Kim et al. reported an improved zT (~1.85) in PbTe-PbSe composite due to the synergistic effect of reduced $\kappa_{ph}$ and enhanced $\alpha^2\sigma$.[21] A notable reduction in $\kappa_{ph}$ was also observed in several composites with nanostructured secondary phase and is mainly attributed to the quantum size effects.[16,22–26] In composite materials, the interface thermal resistance ($R_{int}$) which originates from the acoustic impedance mismatch (AIM) between the phases are rarely considered for optimization of $\kappa_{ph}$ that results in improved zT.[27] However, $R_{int}$ is often used to explain the heat transport mechanism in ceramic and polymer composites, including ZnS/diamond[27], SiC/Al[28], glass/epoxy.[29] These reports alongwith our previous studies on composite materials demonstrate that $R_{int}$ between the phases of the composite is a crucial parameter to design TE composite materials with desired $\kappa_{ph}$.[14,30]

GeTe based materials are promising for TE application in the mid-temperature range (500-800 K). However, pristine GeTe suffers from the intrinsic Ge vacancies that result in a high hole carrier concentration (~$10^{21}$ cm$^{-3}$), high thermal conductivity (~8 W·m$^{-1}$·K$^{-1}$), and low Seebeck coefficient (~30 µV·K$^{-1}$) and hence poor $zT$.[31] Our previous report shows that the simultaneous effect of engineering band structure and lattice dynamics improves $zT$ in Mn-Sb co-doped GeTe at 773 K.[32] Herein, we demonstrate a composite system considering the attuned electronic structure and mismatched phonon structure (AES-MPS) concept to improve zT for GeTe based system. For this purpose, tungsten carbide (WC) as a second phase is used to prepare composite material with $Ge_{0.87}Mn_{0.05}Sb_{0.08}Te$. WC possesses higher electrical and thermal conductivity than $Ge_{0.87}Mn_{0.05}Sb_{0.08}Te$. The large difference in elastic properties between $Ge_{0.87}Mn_{0.05}Sb_{0.08}Te$ and WC has been used to control $\kappa_{ph}$ in the composite using $R_{int}$ between the phases, estimated using AIM and the Debye model. In particular, the effect of WC on electrical transport has been analyzed using the work function of both the materials, measured by the Kelvin probe force microscopy technique. The electronic band structure for the Mn-Sb co-doped GeTe/WC composite is calculated using the density functional theory (DFT). The $\kappa_{ph}$ in the composite is analyzed using the Bruggeman asymmetrical model, which considers $R_{int}$ between the phases. Furthermore, the phonon dispersion calculation for the Mn-Sb doped GeTe/WC composite is also performed to establish the decrease in $\kappa_{ph}$.

## 2. Results and Discussion

### 2.1. Structural Characterization:

X-ray diffraction pattern of the $(1-z)Ge_{0.87}Mn_{0.05}Sb_{0.08}Te/(z)WC$ composite is shown in Fig. 1. $Ge_{0.87}Mn_{0.05}Sb_{0.08}Te$ shows a rhombohedral structure at 300 K, having the following lattice parameters in a hexagonal setting: a=b=4.1709Å, c=10.5612Å. The diffraction intensity of WC is not prominently observed due to its lower volume fraction in the composite. However, the maximum intensity peak of WC is observed in the log-scale (inset of Fig. 1) and confirms its presence in the composite. It is seen that the intensity corresponding to the WC phase increases with the increase in the WC volume fraction ($z$) in the composite.

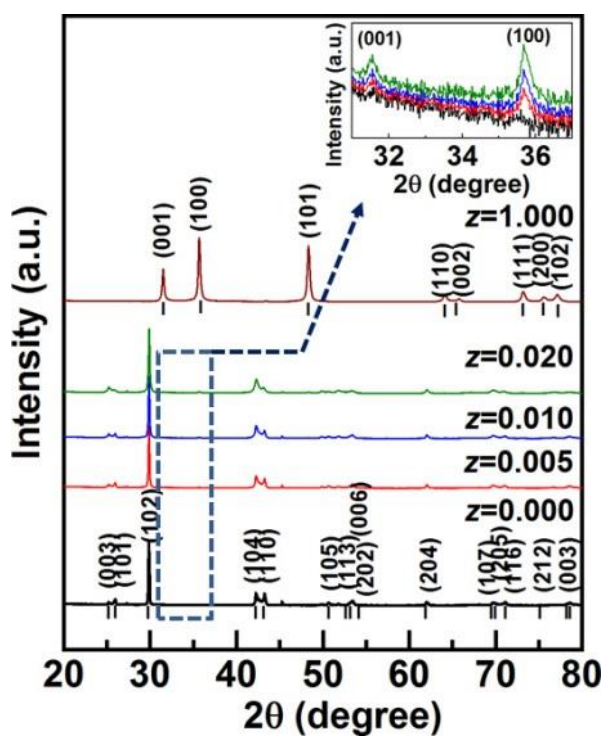

Figure 1: X-ray diffraction pattern for $(1-z)Ge_{0.87}Mn_{0.05}Sb_{0.08}Te/(z)WC$ composite. Inset shows the zoom-in image in log-scale depicting the presence of the diffraction pattern due to WC. The Miller indices and Bragg's position for $Ge_{0.87}Mn_{0.05}Sb_{0.08}Te$ and WC are marked.

Fig. 2(a-d) shows the scanning electron microscope (SEM) image for the polished surface of the $(1-z)Ge_{0.87}Mn_{0.05}Sb_{0.08}Te/(z)WC$ composite. Fig. 2(a) depicts the SEM image of $Ge_{0.87}Mn_{0.05}Sb_{0.08}Te$ sintered pellet which shows the homogeneous nature of the sample. The SEM image of WC powder, used for preparing the composites, is shown in Fig. 2(b). It is observed

that the particle sizes of the WC lie in the range of 150-200 nm. In the composite samples, spherical WC particles are surrounded by $Ge_{0.87}Mn_{0.05}Sb_{0.08}Te$ grains (Fig.2(c-d)). The WC particles are uniformly distributed at the grain boundary, and their concentration increases with the increase in the volume fraction (*z*) in the composite. The bulk density of the sintered pellet is measured using sample mass and its geometrical volume. The relative density of all the samples lies in the range of 98%-99%. Fig. 2(e) shows the energy dispersive x-ray (EDX) spectra of the composite sample with *z*=0.010. The zoom-in image for the same sample is shown in the inset of Fig. 2(e). The presence of WC and other constituents of GeTe is confirmed by the EDX spectrum and is presented in Fig. 2(e). Hence the structural and microstructural analysis confirms the presence of individual phases in the composite samples.

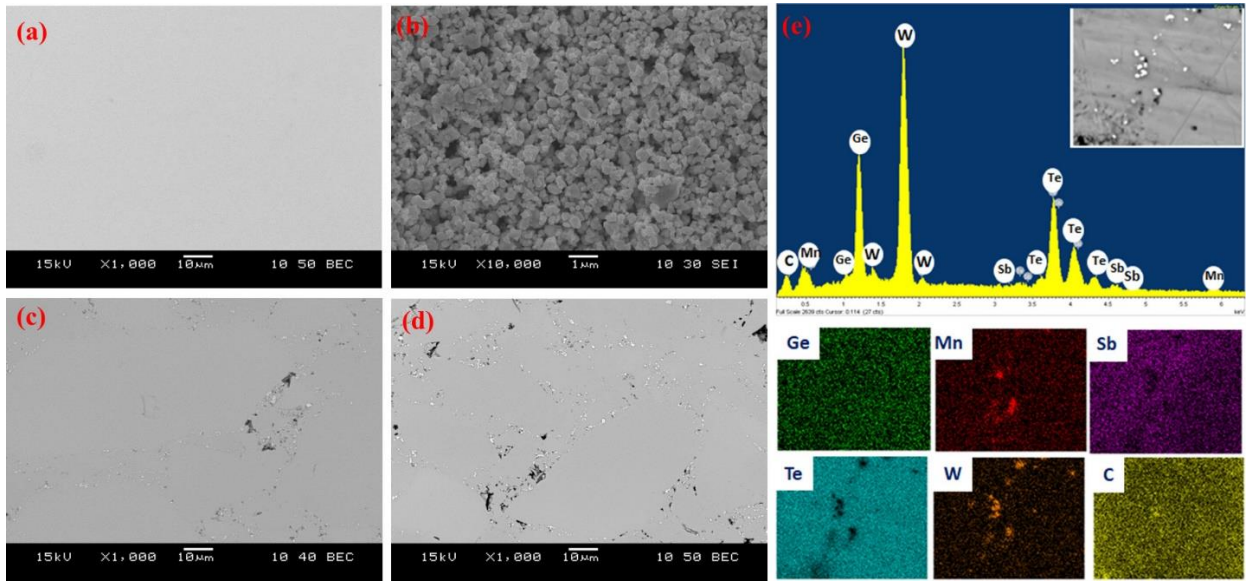

Figure 2: Scanning electron microscope (SEM) images for (a) $Ge_{0.87}Mn_{0.05}Sb_{0.08}Te$ polished pellet (b) WC powder, and (1-*z*)$Ge_{0.87}Mn_{0.05}Sb_{0.08}Te$/(*z*)WC composite with (c) *z*=0.010, (d) *z*=0.020 and (e) EDX spectra for *z*=0.010 is shown. Inset shows the zoom-in image for *z*=0.010. The corresponding elemental mapping for each element present in the sample is also shown.

## 2.2. Electrical Transport Properties:

The electrical conductivity (σ) as a function of temperature for (1-*z*)$Ge_{0.87}Mn_{0.05}Sb_{0.08}Te$/(*z*)WC (0≤*z*≤0.02) is shown in Fig. 3(a). The σ for all the samples decreases with an increase in temperature showing the degenerate semiconducting behavior. Also, we have observed that σ enhances with an increase in the WC volume fraction in the composite. The σ of Ge$_{0.87}$Mn$_{0.05}$Sb$_{0.08}$Te at 300 K is 1150 S·cm$^{-1}$, and it increases to 1342 S·cm$^{-1}$ for *z*=0.010 and 1501 S·cm$^{-1}$ for *z*=0.020. This increase in σ may be attributed to the high σ of the dispersed phase (~50,000 S·cm$^{-1}$ for WC).[33] The literature reports suggest that the conducting dispersed phase at the grain boundary in the composite promotes electrical connectivity to enhance electronic transport.[34]

The electrical conductivity (σ) as a function of WC volume fraction at 300 K is shown in Fig.3(b). Due to microstructure features of the composite it is expected that the current will be flowing through the grain boundaries, where the amount of highly conductive WC phase is relatively high. The theoretical values of electrical conductivity of the composite are calculated using the percolation model [35], given by:

$$\sigma = \sigma_m \left(\frac{z_c - z}{z_c}\right)^{-s} \qquad (2)$$

Here σ$_m$ and σ represent the electrical conductivity of the matrix and the composite respectively, z is the volume fraction, s is a constant, with a well established and universal value of 0.87.[35] This model was fitted to the experimental data using Origin software with percolation threshold as a parameter. Line in the Fig.3(b) represents obtained values of electrical conductivity for (1-z)Ge$_{0.87}$Mn$_{0.05}$Sb$_{0.08}$Te/(z)WC composite with fitted percolation threshold (z$_c$) value of 0.073 (Pearsosn's correlation coefficient R$^2$=0.971). This indicates that the volume fraction of WC required for creation of continuous percolation paths in the composite is relatively small (~7%). The volume fraction used in the present study (z=0.005-0.02) are enough to cause a noticeable increase of composite electrical conductivity.

Next, we discuss the change in carrier concentration (*n*) and carrier mobility (*μ*) with increase in WC volume fraction (z) in the composite using the relation σ=*neμ*, where *e* is electronic charge.[36] The carrier concentration (*n*) measured using Hall measurement and corresponding carrier mobility (μ) estimated using the σ and *n* are shown in Fig.3(c). It is found that the μ of the composite increases with an increase in the WC volume fraction. In general, the addition of the second phase creates scattering centers for charge carriers in bulk semiconducting materials and hence reduces μ. However, a significant enhancement in σ is obtained in the present study and is attributed to the enhanced μ. A similar observation was demonstrated by Zhou et al. in Ag added skutterudites.[34] The increase in μ in the composite sample may be attributed to the filtering of high energy carriers at the interface between these two phases in the composite (discussed later).[37,38]

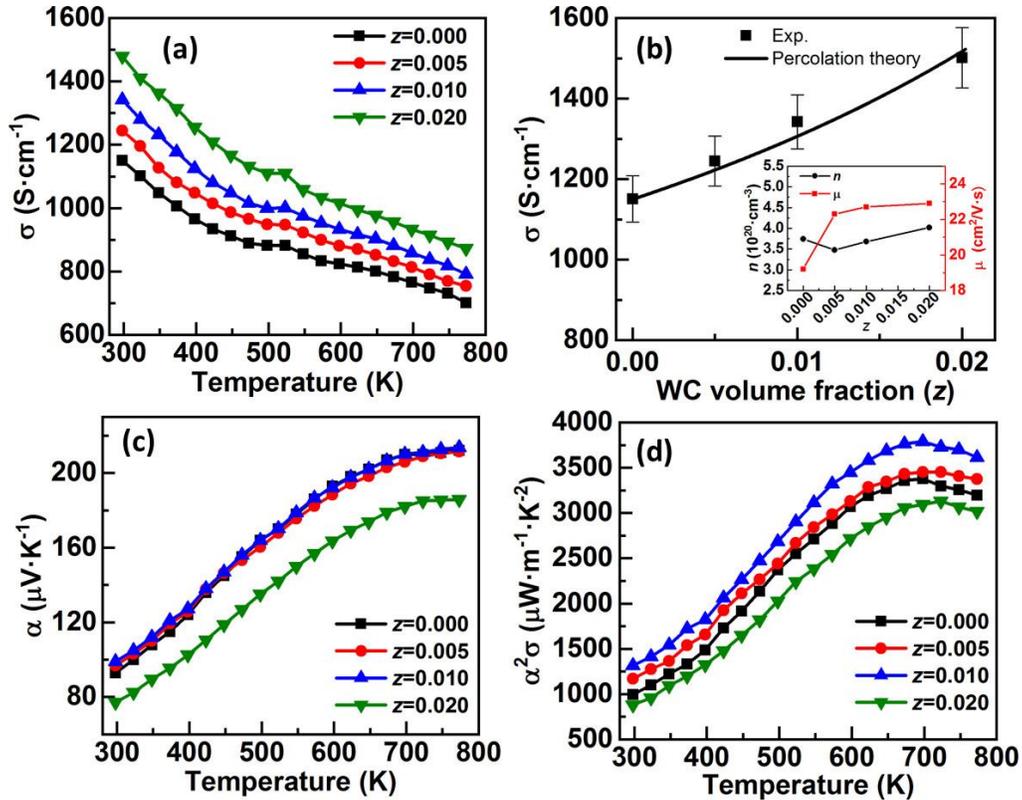

Figure 3: (a) Electrical conductivity (σ) as a function of temperature for (1-$z$)Ge$_{0.87}$Mn$_{0.05}$Sb$_{0.08}$Te/($z$)WC composite. (b) σ as a function of WC volume fraction ($z$) at 300 K. The σ calculated using percolation theory is shown using a solid line. Inset shows the changes in carrier concentration ($n$) and carrier mobility (μ) as a function of $z$. (c) Temperature-dependent Seebeck coefficient (α) and (d) power factor (α$^2$σ) for (1-$z$)Ge$_{0.87}$Mn$_{0.05}$Sb$_{0.08}$Te/($z$)WC composite.

Temperature-dependent Seebeck coefficient (α) for (1-$z$)Ge$_{0.87}$Mn$_{0.05}$Sb$_{0.08}$Te/($z$)WC (0≤$z$≤0.02) is shown in Fig.3(c). The α for all samples increases with an increase in temperature and is consistent with the changes in σ. Although the addition of WC improves the σ significantly, it also enhances α of the composite sample up to $z$=0.010. The α for Ge$_{0.87}$Mn$_{0.05}$Sb$_{0.08}$Te at 300 K is ~93 μV·K$^{-1}$ and increases to (~100 μV·K$^{-1}$) for the composite samples up to $z$=0.010. However, with further increase in WC volume fraction, α decreases to 78 μV·K$^{-1}$ for $z$=0.020. These values of α are consistent with the α$_{av}$ obtained from the STM analysis. The increase and decrease in α are consistent with the carrier concentration change in the composite (Fig.3(b)). The carrier concentration decreases up to $z=0.010$ and then increases. Such an increase in α for small WC volume fraction can be attributed to the carrier energy filtering effect.[37,38] Li et al. showed that the addition of SiC enhances α for PbTe based composite.[22] A similar observation was reported in $Ge_{0.94}Bi_{0.06}Te$/SiC composite.[26] The simultaneous increase in σ and α results in an enhanced power factor ($α^2σ$), as shown in Fig.3(d). The $α^2σ$ increases with the increase in temperature up to ~700 K and then decreases. The power factor increases from 1320 µW·m$^{-1}$·K$^{-2}$ at 300 K and reaches ~3800 µW·m$^{-1}$·K$^{-2}$ at 700 K for the composite with $z=0.010$. Due to a significant decrease in α for higher WC volume fraction ($z=0.020$), the power factor reduces across the temperature range.

### 2.2.1. Kelvin probe force microscopy (KPFM) measurement:

It is noted that the increase in $α^2σ$ for composite is owing to a significant increase in σ. Hence a better understanding of how σ increases in the composite is required. For this purpose, we have investigated the potential barrier at the interface between the $Ge_{0.87}Mn_{0.05}Sb_{0.08}Te$ and WC by measuring the work function for both the individual phases using the Kelvin probe force microscopy (KPFM) technique.[39] KPFM is a vital tool to understand the relative position of the Fermi level in solids.[40] The work function for individual phases is calculated by measuring the contact potential difference (CPD) using the KPFM. The CPD is defined as:

$$\mathrm{CPD} = \frac{\phi_{sample} - \phi_{tip}}{e} \qquad (3)$$

where $\phi_{tip}$ and $\phi_{sample}$ are the work functions of the atomic force microscopy (AFM) probe and sample, respectively. The work function of the AFM probe ($\phi_{tip}$ =4.15 eV) was determined by measuring the potential map of freshly cleaved HOPG with the known value of work function ($\phi_{HOPG}$ =4.6 eV).[41] Further, the work functions of the samples ($\phi_{sample}$) are found using $\phi_{sample}$ = $\phi_{tip}$ + CPD. The surfaces of both the samples ($Ge_{0.87}Mn_{0.05}Sb_{0.08}Te$ and WC) and HOPG were scanned alternatively to determine the CPD values.

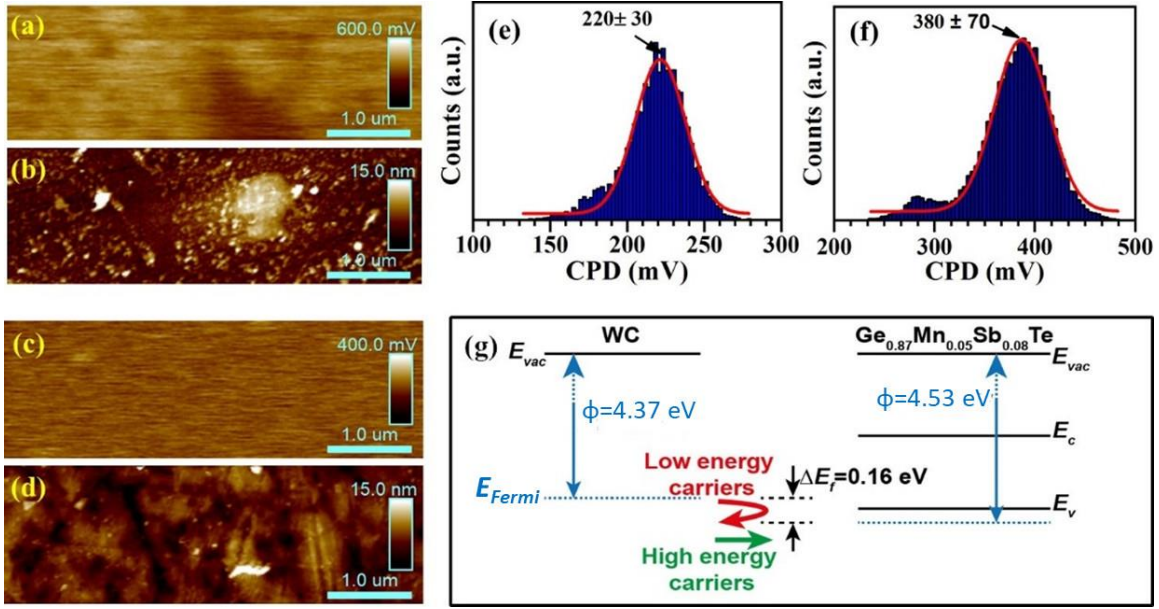

Figure 4: The spatial variation of contact potential difference (CPD) and corresponding 2-dimensional surface topography for (a,b) $Ge_{0.87}Mn_{0.05}Sb_{0.08}Te$, and (c,d) WC. The CPD histogram for (e) WC and, (f) $Ge_{0.87}Mn_{0.05}Sb_{0.08}Te$. (g) The band diagram was estimated from the work function obtained from the Kelvin probe force microscopy (KFPM) for $Ge_{0.87}Mn_{0.05}Sb_{0.08}Te$ and WC. The small Fermi energy difference indicates a crossover of high-energy carriers at the interface.

The spatial variation of CPD and two-dimensional surface topography observed for both phases are shown in Fig.4(a-d). The CPD histograms for $Ge_{0.87}Mn_{0.05}Sb_{0.08}Te$ and WC phases are shown in Fig.4(e-f). The work function obtained from the KPFM measurement for $Ge_{0.87}Mn_{0.05}Sb_{0.08}Te$ and WC is 4.53 eV and 4.37 eV, respectively. It is worth noting that the mean values of CPD for HOPG analyzed before and after measurements are practically the same (449 mV vs. 461 mV), excluding tip contamination. The work function obtained from the KPFM measurement is used to design the band diagram for both the phases and is shown in Fig.4(g). It is noted that for the semiconductor ($Ge_{0.87}Mn_{0.05}Sb_{0.08}Te$) - metal (WC) junction, there can be either Schottky contact (work function of the metal is greater than semiconductor) or Ohmic contact (work function of semiconductor is greater than metal).[42,43] The present study shows that the work function for WC is smaller than $Ge_{0.87}Mn_{0.05}Sb_{0.08}Te$. It indicates that charge carriers can flow from WC to $Ge_{0.87}Mn_{0.05}Sb_{0.08}Te$, supporting the enhanced σ in the system. Also, the energy difference (ΔE$_f$) between these two materials is quite small (0.16 eV) that helps to enhance σ in the composite. This small difference in ΔE$_f$ may scatter the lower energy carrier at the interface and allows the high energy carriers with increased µ to improve α due to energy filtering.[25,44] This indicates that the small mismatch in ΔE$_f$ enhances both σ and α in the composite.

## 2.3: Electronic Structure and Work function Calculations:

To further explore the enhancement of electrical conductivity in Ge$_{19}$MnSb$_2$Te$_{24}$/WC composite, we have employed density functional theory (DFT). Firstly, we have examined the stability of Ge$_{19}$MnSb$_2$Te$_{24}$/WC composite by calculating the binding energies of all systems, which is defined as:

$$E_b = E(\text{Ge}_{19}\text{MnSb}_2\text{Te}_{24}/\text{WC}) - E(\text{Ge}_{19}\text{MnSb}_2\text{Te}_{24}) - E(\text{WC})$$

where $E$(Ge$_{19}$MnSb$_2$Te$_{24}$/WC), $E$(Ge$_{19}$MnSb$_2$Te$_{24}$) and $E$(WC) are respectively the total energies of Ge$_{19}$MnSb$_2$Te$_{24}$/WC composite, Ge$_{19}$MnSb$_2$Te$_{24}$ matrix and WC. Fig. 5 (a-c) shows the optimized geometries of Ge$_{19}$MnSb$_2$Te$_{24}$, WC, and Ge$_{19}$MnSb$_2$Te$_{24}$/WC composite, respectively.

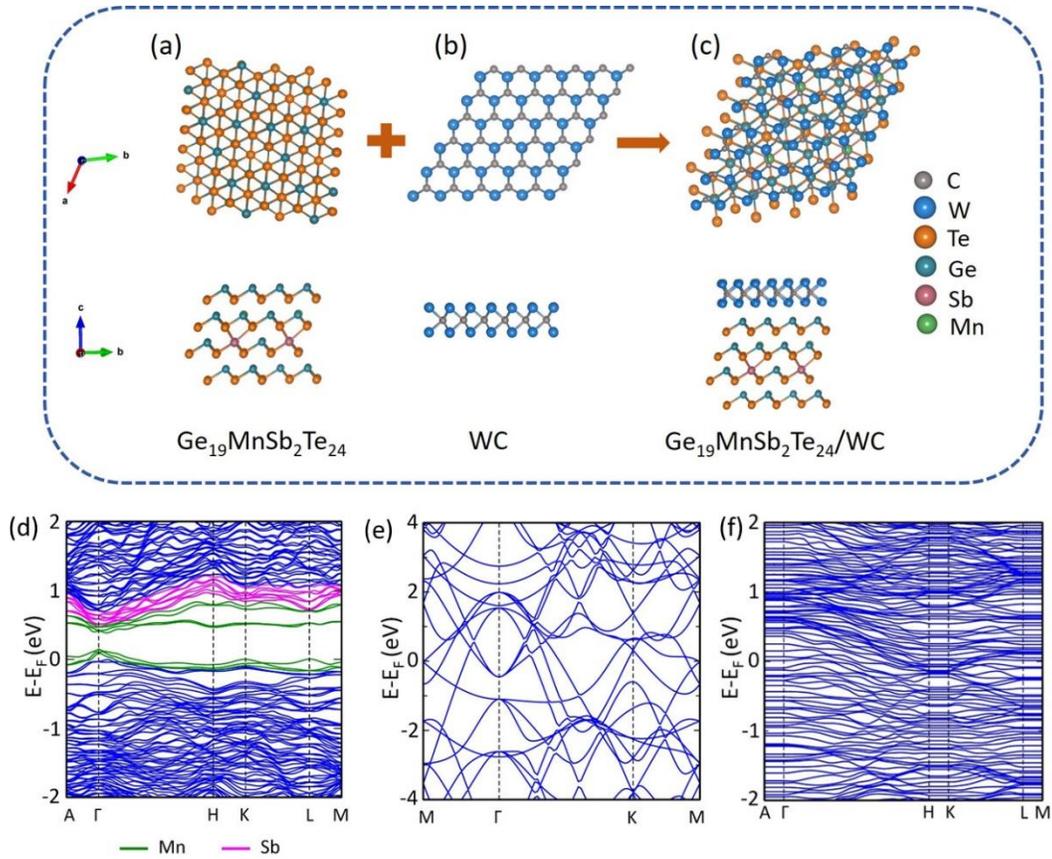

Figure 5: Side and top views of optimized geometries of (a) Ge$_{19}$MnSb$_2$Te$_{24}$, (b) WC, and (c) Ge$_{19}$MnSb$_2$Te$_{24}$/WC composite. Electronic band structures of (d) Ge$_{19}$MnSb$_2$Te$_{24}$, (e) WC and (f) Ge$_{19}$MnSb$_2$Te$_{24}$/WC composite.

The binding energy of $Ge_{19}MnSb_2Te_{24}$/WC composite comes out to be -1.85 eV. The small and negative value of binding energy implies that the composite is thermodynamically stable. Subsequently, we have calculated the band structures for $Ge_{19}MnSb_2Te_{24}$, WC and $Ge_{19}MnSb_2Te_{24}$/WC (see Fig. 5(d-f)). As we can see from Fig. 5(e), there is no gap in the band structure, which indicates the metallic nature of WC. On the other hand, $Ge_{19}MnSb_2Te_{24}$ is a *p*-type semiconductor. This leads to the possibility of transfer of electrons from WC to $Ge_{19}MnSb_2Te_{24}$ matrix in $Ge_{19}MnSb_2Te_{24}$/WC composite.

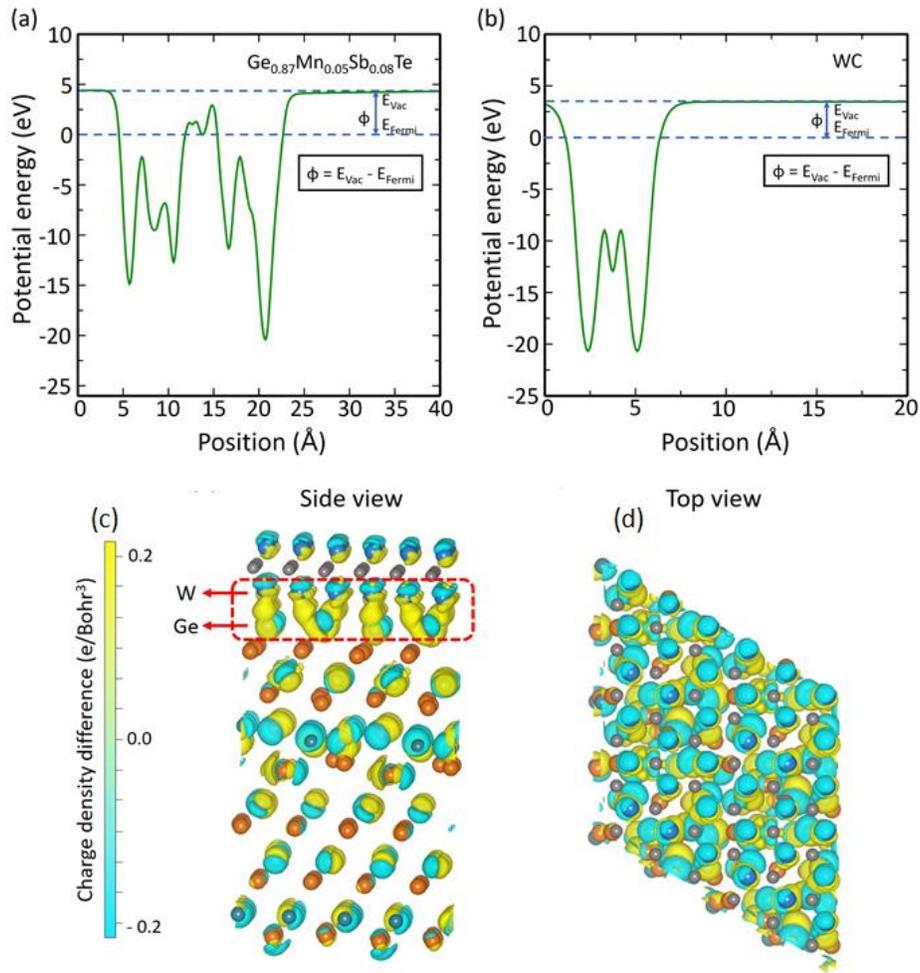

Figure 6: Electrostatic potential energy of (a) $Ge_{19}MnSb_2Te_{24}$ and (b) WC along Z-direction. (c) Side and (d) Top view of charge density difference plot for $Ge_{19}MnSb_2Te_{24}$/WC composite, where the cyan and yellow fields represent the electron-accumulation and depletion, respectively.

In order to examine the charge transfer between WC and $Ge_{19}MnSb_2Te_{24}$, we have plotted the electrostatic potential energy for both systems. Fig. 6(a-b) shows the electrostatic potential energy of $Ge_{19}MnSb_2Te_{24}$ and WC along Z-direction. From these calculations, the obtained work functions for $Ge_{19}MnSb_2Te_{24}$ and WC are 4.51 eV and 4.36 eV, respectively, which are in well agreement with the experimental results. The smaller work function of WC indicates that charge can flow from WC to $Ge_{19}MnSb_2Te_{24}$, leading to increase in the electrical conductivity. To better understand the charge transfer, we have plotted the 3D charge density difference of the composite as shown in Fig. 6(c-d). The yellow and cyan fields represent the electron-accumulation and depletion, respectively. In Fig. 6(c), it can be clearly seen that W atoms transfer the electrons to Ge atoms (represented by yellow color). This charge transfer enhances the electrical conductivity of the $Ge_{19}MnSb_2Te_{24}$/WC composite.

## 2.4. Thermal Transport Properties:

Temperature-dependent total thermal conductivity (κ) for $(1-z)Ge_{0.87}Mn_{0.05}Sb_{0.08}Te/(z)WC$ composite is shown in Fig.7(a). The κ decreases with an increase in temperature for all the samples. However, it increases with the increase in WC volume fraction in composite up to 500 K and decreases with further increase in temperature for composite samples up to $z=0.010$. A noticeable increase in κ at all temperatures is seen for the sample with $z=0.020$. This increase in κ is logical since WC possesses very high κ (~170 W·m$^{-1}$·K$^{-1}$) at 300 K. However, it does not follow the rule of mixture despite a significant increase in σ. Since κ consists of electronic thermal conductivity ($κ_e$) and phonon thermal conductivity ($κ_{ph}$) components, i.e., $κ = κ_e + κ_{ph}$, the addition of WC volume fraction in the composite improves σ and hence must increase $κ_e$. To elucidate the nature of electron and phonon contribution to thermal conductivity, both $κ_e$ and $κ_{ph}$ in the composite system are separated from κ by calculating the $κ_e$ using Wiedemann Franz law $κ_e = LσT$, where the Lorenz number (L) was calculated using the Snyder equation: $L = \left(1.5 + \exp\left[-\frac{|\alpha|}{116}\right]\right)$ ×10$^{-8}$ W·Ω·K$^{-2}$ with α in μV·K$^{-1}$. [45]

### 2.4.1. Phonon thermal conductivity and Bruggeman's model:

Temperature-dependent $\kappa_{ph}$ for (1-z)Ge$_{0.87}$Mn$_{0.05}$Sb$_{0.08}$Te/(z)WC composite is shown in Fig.7(b). It is worth noting that $\kappa_{ph}$ of the composite decreases with the increase in temperature, and interestingly, it also decreases with the increase in WC volume fraction. The changes in electronic ($\kappa_e$) and phonon thermal conductivity ($\kappa_{ph}$) with an increase in WC volume fraction at 300 K are shown in the inset of Fig.7(b). It is noted that $\kappa_{ph}$ of WC is relatively higher (~135 W·m$^{-1}$·K$^{-1}$ at 300 K) as compared to Ge$_{0.87}$Mn$_{0.05}$Sb$_{0.08}$Te. However, $\kappa_{ph}$ of the composite decreases with the addition of the WC volume fraction. This decrease in $\kappa_{ph}$ for composite is interesting and is analyzed using Bruggeman's asymmetrical model that considers $R_{int}$ between the phases in the composite. The $R_{int}$ for the composite is estimated using the AIM and the Debye model.[46,47] The AIM model shows that the phonon propagating from one material to another can be reflected if there is a mismatch in the acoustic impedance (Z=v×ρ) between the two materials, where $v$ is the sound velocity, and ρ is the sample density.[30] The sound velocity ($v$) and sample density (ρ) measured for both Ge$_{0.87}$Mn$_{0.05}$Sb$_{0.08}$Te and WC and their calculated acoustic impedance are presented in Table I. The probability of phonon transmission ($\eta = pq$) at the interface between Ge$_{0.87}$Mn$_{0.05}$Sb$_{0.08}$Te (matrix) and WC (dispersed phase) is 0.0436, where $p = \frac{4Z_A Z_B}{(Z_A + Z_B)^2}$, $q = \frac{1}{2}\left(\frac{v_m}{v_d}\right)^2$ are calculated using the sound velocity of the matrix ($v_m$) and dispersed phase ($v_d$).[48] The acoustic impedance for the matrix is $Z_A$ and that of the dispersed phase is $Z_B$.

Table I: The measured values of sample density (ρ), sound velocity ($v$) used in acoustic impedance mismatch (AIM) model to calculate the acoustic impedance (Z) and Transmission coefficient ($p$) of phonons for Ge$_{0.87}$Mn$_{0.05}$Sb$_{0.08}$Te and WC samples.

| Sample Name | ρ (g·cm$^{-3}$) | $v$ (m·sec$^{-1}$) | | Z (kg·m$^{-2}$s$^{-1}$) | | $p_{av}$ |
| --- | --- | --- | --- | --- | --- | --- |
| | | Trans. ($v_l$) | Long. ($v_t$) | Trans. | Long. | (%) |
| Ge$_{0.87}$Mn$_{0.05}$Sb$_{0.08}$Te | 5.74 | 1870 | 3220 | 10734 | 18483 | 48.3 |
| WC | 15.43 | 4400 | 7180 | 67892 | 110787 | |

The probability of phonon transmission obtained from the AIM model is used to calculate the $R_{int}$ between the Ge$_{0.87}$Mn$_{0.05}$Sb$_{0.08}$Te/WC phases following the Debye model: $R_{int} = \frac{4}{\rho c_p v_D \eta}$, where $c_p$ is the specific heat capacity of the matrix, $v_D = \left(\frac{3}{\frac{1}{v_l^3} + \frac{2}{v_t^3}}\right)^{1/3}$ is the Debye velocity. Using the values of $c_p$, $\rho$, $v_D$, and $\eta$, the estimated $R_{int}$ for Ge$_{0.87}$Mn$_{0.05}$Sb$_{0.08}$Te and WC is 2.535× 10$^{-6}$ m$^2$·K·W$^{-1}$. This value of $R_{int}$ is higher than several TE composites and is beneficial for improving phonon scattering between the composite phases.[14,30] The $R_{int}$ between the phases gives a critical grain size called Kapitza radius ($a_K = R_{int} \cdot \kappa_{ph,m}$) below which the $\kappa_{ph}$ of the composite reduces.[27] Using the phonon thermal conductivity of the matrix ($\kappa_{ph,m}$) and $R_{int}$, $a_K$ for Ge$_{0.87}$Mn$_{0.05}$Sb$_{0.08}$Te/WC composite is 3.40 μm at 300 K. A high $R_{int}$ provides a larger $a_K$ and it suggests that $\kappa_{ph}$ of the composite can be reduced if the particle size of the dispersed phase is smaller than 3.40 μm. The consideration of AIM and the Debye model is important to estimate the critical size for a composite to reduce its $\kappa_{ph}$.

Further, Bruggeman's asymmetrical model is used to analyze the $\kappa_{ph}$ of the Ge$_{0.87}$Mn$_{0.05}$Sb$_{0.08}$Te/WC composite, using $a_K$ obtained for the composite, by the formula:[49]

$$(1-z)^3 = \left(\frac{\kappa_{ph,m}}{\kappa_{ph}}\right)^{\frac{1+2s}{1-\alpha}} \left(\frac{\kappa_{ph} - \kappa_{ph,d}(1-s)}{\kappa_{ph,m} - \kappa_{ph,d}(1-s)}\right)^{\frac{3}{1-s}} \tag{4}$$

where $\kappa_{ph,d}$ and $\kappa_{ph}$ are the thermal conductivity of the dispersed phase (WC) and the composite, respectively, and $s = a/a_K$ where $a$ is the particle size of the dispersed phase. $\kappa_{ph}$ calculated using the Bruggeman's asymmetrical model for Ge$_{0.87}$Mn$_{0.05}$Sb$_{0.08}$Te/WC composite with $z$=0.020 at 300 K as a function of different particle size of WC is shown in the inset of Fig.7(c). As can be seen from the curve, $\kappa_{ph}$ decreases with the decrease in the particle size of WC. It reduces below $\kappa_{ph,m}$ when $a<a_K$ (marked by the dotted line in the inset of Fig.7(c)).

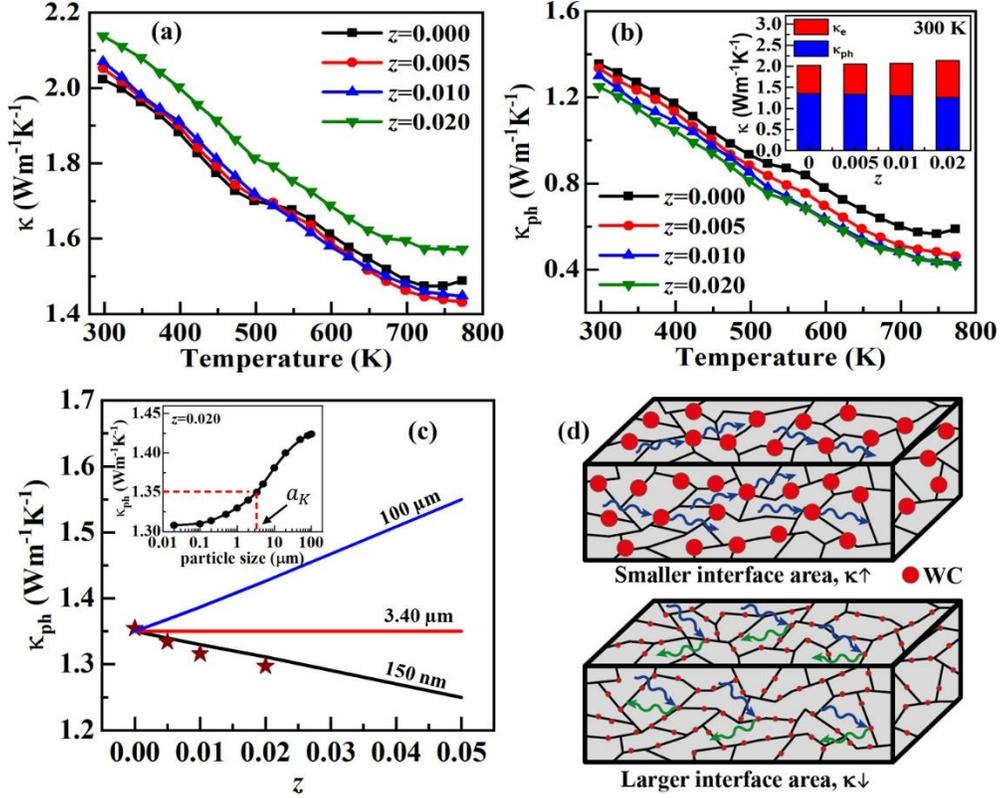

Figure 7: (a) Total thermal conductivity ($\kappa$), (b) phonon thermal conductivity ($\kappa_{ph}$) as a function of temperature for $(1-z)Ge_{0.87}Mn_{0.05}Sb_{0.08}Te/(z)WC$ composite. Inset of (b) shows the contribution of $\kappa_e$ and $\kappa_{ph}$ at 300 K, (c) The $\kappa_{ph}$ calculated (solid lines) as a function of WC volume fraction for $(1-z)Ge_{0.87}Mn_{0.05}Sb_{0.08}Te/(z)WC$ for the different particle size of WC, using Bruggeman's asymmetrical model. The symbols represent the $\kappa_{ph}$ obtained from the total $\kappa$. Inset of (c) shows the $\kappa_{ph}$ calculated for composite with $z=0.020$ as a function of WC particle size. (d) A schematic of composite materials with different particle sizes of WC (red circles) is shown. The phonons are reflected for smaller WC particle size due to the prominent effect of interface thermal resistance between phases due to increased interfacial area/surface to volume ratio.

Figure 7(c) depicts $\kappa_{ph}$ as a function of WC volume fraction ($z$) for the different particle size of WC at 300 K. The $\kappa_{ph}$ increases when $a>a_K$ ~3.40 μm. It indicates that even a higher acoustic mismatch between $Ge_{0.87}Mn_{0.05}Sb_{0.08}Te$ and WC increases $\kappa_{ph}$ when $a>a_K$. However, $\kappa_{ph}$ remains the same for $a\sim a_K$. It decreases when $a<a_K$ and indicates that $R_{int}$ becomes prominent for $a<a_K$ and hence reduces $\kappa_{ph}$. It also suggests a strong correlation between $R_{int}$ and $a_K$.[14] In other words, the smaller particle size of the dispersed phase enhances the surface to volume ratio and hence prominently reduces $\kappa_{ph}$. A schematic of $Ge_{0.87}Mn_{0.05}Sb_{0.08}Te$/WC composite for the different particle sizes of WC is shown in Fig.7(d). It shows that the larger particle size of the dispersed phase possesses a lower interface area, which leads to an increase in $\kappa_{ph}$ even if $R_{int}$ is high. On the other hand, if the particle size of the dispersed phase is smaller than $a_K$, the $R_{int}$ becomes prominent with increased interface area/surface to volume ratio and reduces $κ_{ph}$.

### 2.4.2: Phonon Dispersion Calculation:

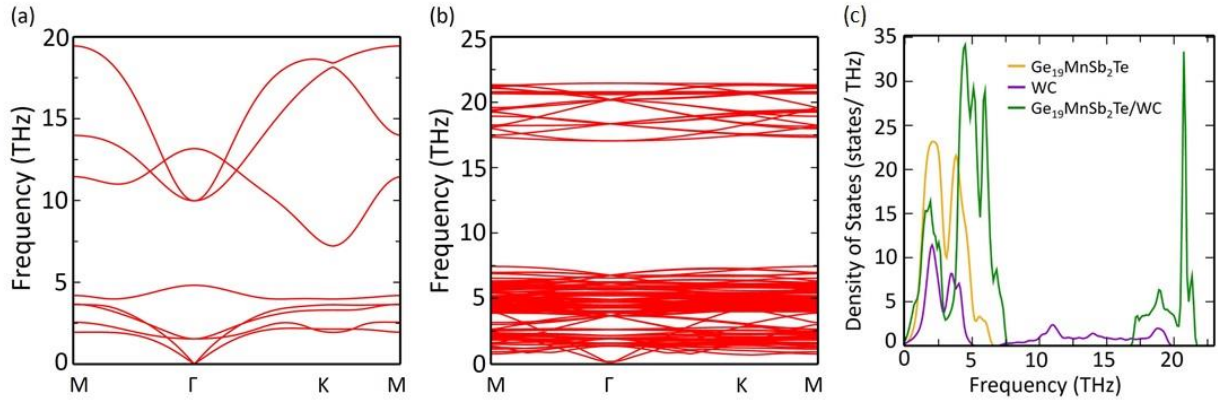

Figure 8: Phonon dispersion curves for (a) WC (b) Ge$_{19}$MnSb$_2$Te$_{24}$/WC composite. (c) Phonon density of states for Ge$_{19}$MnSb$_2$Te$_{24}$, WC and Ge$_{19}$MnSb$_2$Te$_{24}$/WC composite.

Next, to examine the thermal stability, phonon dispersion curves for WC and Ge$_{19}$MnSb$_2$Te$_{24}$/WC were plotted using phonopy code (see Fig. 8 (a-b)). Also, as shown in Fig. 8 (c), the phonon density of states for Ge$_{19}$MnSb$_2$Te$_{24}$, WC and Ge$_{19}$MnSb$_2$Te$_{24}$/WC composite are plotted which shows their contribution in different frequency range. In general, the phonon dispersion is indicated by the $ω$ vs $k$ plot, and the gradient of $ω$ vs $k$ curve gives the $v_g$ (phonon group velocity), where $v_g$ = d$ω$/d$k$. As can be seen in Fig. 8 (a) and (b), the gradient of phonon curve for Ge$_{19}$MnSb$_2$Te$_{24}$/WC is lower than that of WC and Ge$_{19}$MnSb$_2$Te$_{24}$. This suggests that the Ge$_{19}$MnSb$_2$Te$_{24}$/WC has lower lattice thermal conductivity in comparison to Ge$_{19}$MnSb$_2$Te$_{24}$ and WC. These findings are in well agreement the experimental results.

### 2.5. Figure of Merit and Efficiency:

Using the experimentally observed TE parameters: α, σ and κ, the figure of merit ($zT$) of the composite is calculated and is shown in Fig.9(a). The $zT$ increases with an increase in temperature for all the samples. The $zT$ also increases with WC volume fraction and shows a maximum of 1.93 at 773 K for the sample with z=0.010. This enhancement in zT is attributed to the simultaneous rise in σ and α for composite along with reduced $\kappa_{ph}$ owing to the AIM between the phases. However, the higher WC fraction in the composite reduces zT due to a significant reduction in α.

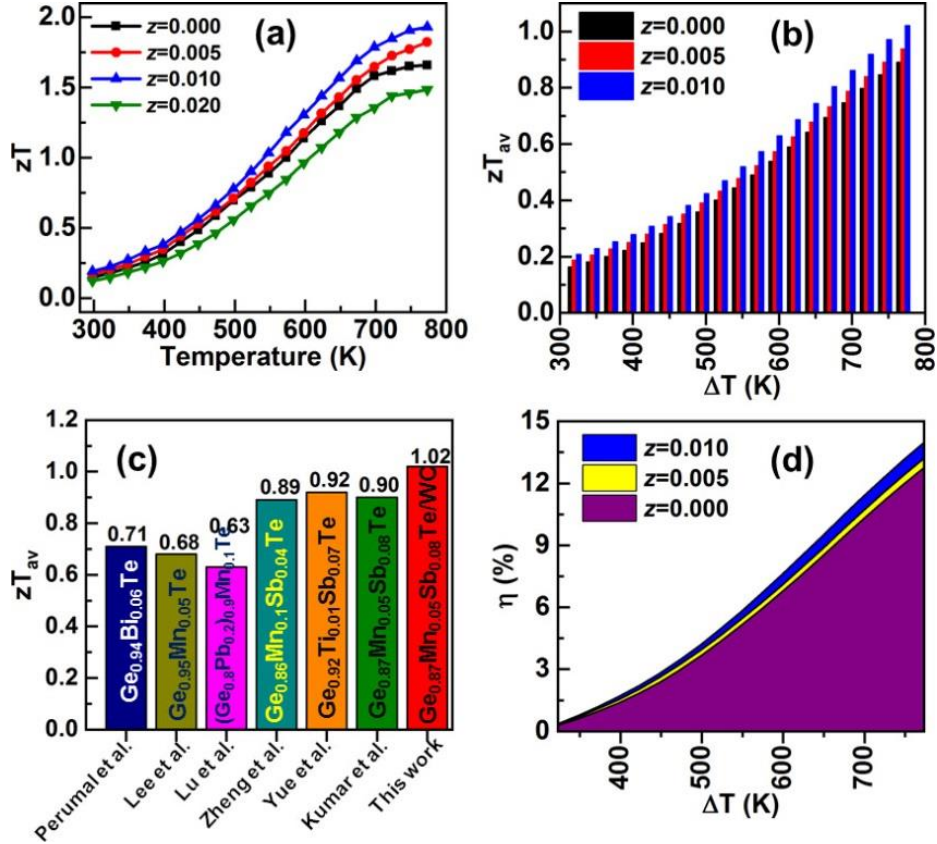

Figure 9. (a) Figure of merit (zT) (b) average zT for the (1-z)Ge$_{0.87}$Mn$_{0.05}$Sb$_{0.08}$Te/(z)WC composite, (c) average zT compared with the values reported in the literature[50–54], (d) The energy conversion efficiency calculated considering similar n-type leg for (1-z)Ge$_{0.87}$Mn$_{0.05}$Sb$_{0.08}$Te/(z)WC composite.

The energy conversion efficiency of TE devices made up of TE materials is defined as $\eta = \frac{\Delta T}{T_H} \frac{\sqrt{1+zT_{av}}-1}{\sqrt{1+zT_{av}}+\frac{T_C}{T_H}}$. This expression suggests that a high $\eta$ requires a high average ($zT_{av} = \frac{1}{T_H-T_C}\int_{T_C}^{T_H} zT dT$) across a wide temperature difference ($\Delta T = T_H - T_C$) between the hot ($T_H$) and cold ($T_C$) temperature ends of the device. The $zT_{av}$ calculated for the (1-z)Ge$_{0.87}$Mn$_{0.05}$Sb$_{0.08}$Te/(z)WC composite with 300 K and 773 K as $T_C$ and cold $T_H$, respectively, is shown in Fig.9(b). The $zT_{av}$ also increases for the sample with z=0.010. It reaches ~1.02 for a temperature difference of 473 K. The obtained $zT_{av}$ in the present study is higher than reported in several promising literature reports.[50–54] A comparison for the same is shown in Fig.9(c). Further, using the $zT_{av}$ obtained in the present study for p-type material and assuming a corresponding similar n-type leg, theoretical energy conversion efficiency (η) is calculated for the temperature difference (ΔT) between the hot and cold end temperature and shown in Fig.9(d). A maximum energy conversion efficiency ~14% is obtained for (1-z)Ge$_{0.87}$Mn$_{0.05}$Sb$_{0.08}$Te/(z)WC for z=0.010 composite and is higher than several promising TE materials reported in the literature.[50–54]

## 3. Conclusion

This study demonstrates the influence of work function and acoustic impedance mismatch (AIM) on electronic and phonon transport properties, respectively, in Ge$_{0.87}$Mn$_{0.05}$Sb$_{0.08}$Te/WC composite. X-ray diffraction analysis confirms the individual phases in the composite, which is further supported by electron microscopy images and energy dispersive x-ray spectroscopy analysis. Enhancement in the composite's electrical conductivity (σ) is attributed to the increase in carrier mobility (μ). It is also analyzed using the work function measurement by the Kelvin probe force microscopy technique. The lower work function of WC compared to Ge$_{0.87}$Mn$_{0.05}$Sb$_{0.08}$Te gives high mobility charge carriers to the system, and hence increases σ. The AIM between the composite phases leads to a high interface thermal resistance ($R_{int}$), beneficial for improving phonon scattering. A correlation between $R_{int}$ and the Kapitza radius ($a_K$) decreases the phonon thermal conductivity ($κ_{ph}$) of the composite, supported by Bruggeman's asymmetrical model. The simultaneous effect of work function and AIM due to WC leads to improved power factor and reduced $κ_{ph}$. A maximum $zT$ of 1.93 at 773 K with a $zT_{av}$~1.02 for a temperature difference of 473 K is obtained for (1-z)Ge$_{0.87}$Mn$_{0.05}$Sb$_{0.08}$Te/(z)WC with z=0.010. It results in the maximum energy conversion efficiency ($\eta$) of ~14%. This study shows promise to develop efficient thermoelectric composite materials further having similar electronic structure and different elastic properties (considering the correlation between $R_{int}$ and $a_K$).

## 4. Experimental Section

### 4.1. Synthesis and Structural Characterization:

The synthesis of $Ge_{0.87}Mn_{0.05}Sb_{0.08}Te$ has been carried out by the direct melting of elements (Ge, Mn, Sb, and Te) with purity >99.99% (Alfa Aesar) in evacuated quartz ampoules following our previous report in GeTe.[32] Further, the $(1-z)Ge_{0.87}Mn_{0.05}Sb_{0.08}Te/(z)WC$ composite with $z$=0.000, 0.005, 0.010, and 0.020 is prepared by mixing the different volume percentage of WC (Sigma Aldrich, 99.9%). The composite mixture was ground to mix homogeneously in a liquid medium (acetone) using a mortar-pestle. The mixture was then sintered using pulsed electric current sintering (PECS) technique in Ar (5N) atmosphere at 873 K for 5 minutes under a uniaxial pressure of 50 MPa with a heating cooling rate of 70 K/min and 50 K/min, respectively. The obtained cylindrical pellets of 10 mm diameter and 12 mm length were cut to proper dimensions using a precise wire saw for further measurements. The surface morphology and chemical analysis of the polished sample surface were done using a scanning electron microscope (SEM, NOVA NANO 200, FEI EUROPE Company) equipped with an EDXS analyzer. The X-ray diffraction of samples was obtained by the D8 ADVANCE (BRUKER) diffractometer using Ni- filtered Cu-K$_\alpha$ radiation ($\lambda$=1.5406 Å).

### 4.2. Electrical and thermal transport properties:

The spatial variation of Seebeck coefficient for the $(1-z)Ge_{0.87}Mn_{0.05}Sb_{0.08}Te/(z)WC$ composite was done using a scanning thermoelectric microprobe (STM) at 300 K with a spatial resolution of 50 μm. Thermal diffusivity for all the samples was measured using the laser flash analysis (LFA-457, NETSCH) apparatus in the Ar (5N) atmosphere (30 ml/min). Specific heat was determined simultaneously with the thermal diffusivity using pyroceram 9606 as reference material. The sample density was measured using sample mass and its geometric volume. Electrical conductivity and Seebeck coefficient were measured under Ar (5N) atmosphere (50 ml/min) by the SBA 458 (NETSCH) apparatus. The uncertainty of the Seebeck coefficient and electrical conductivity measurements is 7% and 5%, respectively. The estimated uncertainty in thermal conductivity is 7%. The carrier concentration was measured at 300 K using physical properties measurement system (PPMS, Quantum Design) under the magnetic field of ±3 T. Work function measurements were performed using atomic force microscope (Dimension ICON, Bruker) working in the Peak Force KPFM mode. PFQNE-AL probes (with a nominal spring constant of 0.8 N/m) were used for capturing of topography and potential maps under ambient conditions. The images were captured with the resolution of 84 x 256 pixels. The work function of AFM tip ($\phi_{tip}$ = 4.15 eV) was determined by measuring the CPD of freshly cleaved HOPG with known value of work function (4.6 eV). The CPDs of HOPG before and after measurements were virtually the same (0.45 V and 0.46 V) excluding tip contamination. The work function of the analyzed sample was calculated using equation 2.

**4.3. Theoretical Methods:**

The density functional theory (DFT) [55][56] calculations were performed using the plane-wave-based pseudopotential approach, as implemented in the Vienna Ab initio Simulation Package (VASP)[57,58]. The self-consistency loop was converged with a total energy threshold of 0.01 meV. The structures were fully relaxed until the Heymann–Feynman forces on each atom were less than $10^{-5}$ eV/Å. The structural optimization was carried out using generalized gradient approximation (GGA) expressed by the Perdew–Burke–Ernzerhof (PBE)[59] exchange-correlation functional. The effects of doping were considered by substituting Mn and Sb atoms at the specific sites of Ge atoms. 2x2x1 and 3x3x1 supercells were employed for GeTe and WC so that there exists a minimum lattice mismatch of ~3.25 %. A 6×6×1 $k$-mesh was used for Brillouin zone sampling for the $Ge_{0.87}Mn_{0.05}Sb_{0.08}Te$/WC composite. The periodic units were separated by a vacuum layer with 20 Å thickness along Z-direction to prevent spurious interactions between periodic images. The two-body vdW interaction as devised by Tkatchenko-Scheffler has been employed.[60,61] The correction parameter is based on Hirshfield partitioning of the electron density. The electron wave function was expanded in a plane-wave basis set with an energy cutoff of 600 eV. Spin–orbit coupling interactions owing to heavy atoms were included when calculating the electronic structures. Phonon calculations were obtained within the harmonic approximation and using a finite displacement method[62]. A 2×2×2 supercell was set for the calculations. The starting parameters for the calculations were the values obtained from the refinement.

**Declaration of Competing Interest**

The authors declare no competing financial interests.

**Acknowledgements**

The authors would like to thank the Foundation for Polish Science for financial support (TEAM-TECH/2016-2/14 grant "New approach for the development of efficient materials for direct conversion of heat into electricity") under the European Regional Development Fund. The beneficiary institution of the grant is Faculty of Materials Science and Ceramics, AGH University of Science and Technology, Kraków, Poland. PB acknowledges UGC, India, for the senior research fellowship [1392/(CSIR-UGC NET JUNE 2018)]. SB acknowledges the financial support from SERB under the core research grant (grant no. CRG/2019/000647). SB and PB thank HPC IIT Delhi for providing computing hours.